\documentclass[sigplan,10pt]{acmart}
\renewcommand\footnotetextcopyrightpermission[1]{}  
\makeatletter
\renewcommand\@formatdoi[1]{\ignorespaces}
\makeatother
\usepackage{url}
\usepackage{caption}
\usepackage{subcaption}
\usepackage{amsmath,amssymb,amsfonts}
\usepackage{algorithmic}
\usepackage{graphicx}
\usepackage{textcomp}
\usepackage{xcolor}
\usepackage{booktabs} 
\usepackage{makecell}
\usepackage{upgreek}
\usepackage{hyperref}
\setcopyright{rightsretained}

\AtBeginDocument{%
  \providecommand\BibTeX{{%
    \normalfont B\kern-0.5em{\scshape i\kern-0.25em b}\kern-0.8em\TeX}}}

\setcopyright{none}

\begin{document}
\pagestyle{plain}

\title{Enabling efficient application monitoring in cloud data centers using SDN}

\author{Mona Elsaadawy}
\email{mona.elsaadawy@mail.mcgill.ca}
\author{Bettina Kemme}
\email{Kemme@cs.mcgill.ca}
\affiliation{%
  \institution{School of Computer Science, McGill University}
  \city{Montreal}
  \state{Quebec}
  \country{Canada}
}
\author{Mohamed Younis}
\email{younis@cs.umbc.edu}
\affiliation{%
  \institution{Department of Computer Science and Electrical Engineering, UMBC}
  \city{Baltimore}
  \state{Maryland}
  \country{USA}
}


\begin{abstract}
Software Defined Networking (SDN) not only enables agility through the realization of part of the network functionality in software but also facilitates offering advanced features at the network layer. Hence, SDN can support a wide range of middleware services; network performance monitoring is an example of these services that are already deployed in practice. In this paper, we exploit the use of SDNs to efficiently provide application monitoring functionality. The recent rise of complex cloud applications has made performance monitoring a major issue. We show that many performance indicators can be inferred from messages exchanged among application components. By analyzing these messages, we argue that the overhead of performance monitoring could be effectively moved from the end hosts into the SDN middleware of the cloud infrastructure which enables more flexible placement of logging functionality. This paper explores several approaches for supporting application monitoring through SDN. In particular, we combine selective forwarding in SDN to enable message filtering and reformatting, and propose a customized port sniffing technique. We describe the implementation of the approach within the standard SDN software, namely OVS. We further provide a comprehensive performance evaluation to analyze advantages and disadvantages of our approach, and highlight the trade-offs.
\end{abstract}




\maketitle

\section{Introduction}\label{intro}

Many application domains have started to move their services into the cloud, e.g.,e-commerce, health management, education, entertainment, and many more. 
Many cloud applications are often deployed as multi-component systems where the individual components and their replicas might run across tens, hundreds or even thousands of nodes and may communicate with other similarly distributed services to accomplish their goals. For example, user-facing components at Google comprise 100s to 1000s of nodes that interact with each other and with other services (e.g., a spell-checking service) to handle user requests. Even simple web applications today consist of multiple distributed tiers that interact with each other and may be integrated with a third party such as Facebook, payment gateways, etc. to accomplish one transaction. 
The performance of these applications has a direct impact on business metrics such as revenue and customer satisfaction. Examples of this impact are well documented. For example, Google loses 20\% traffic for additional delay of 0.5 second to their page-load time and Amazon loses 1\% of revenue for every 100 ms in latency increase \cite{statisticspaper}. 
Given the distributed character of the applications,  effective logging and monitoring  are important to keep track of performance, and to detect, diagnose and resolve performance-related problems.

Apart of standard hardware resource utilization measurements (CPU, memory, etc.), 
application layer metrics such as request service time or characteristics of service call distributions are also highly relevant~\cite{Timothy2016}. Common approaches to collect the relevant data are to instrument the application or the underlying platform to create log messages, such as Apache Tomcat's log valves \cite{Apache}. An obvious disadvantage is the application and/or platform dependence. Additionally, network monitoring middleware tools such as Wireshark \cite{wireshark}, that perform live tracking of message flows and allow for sophisticated analysis of captured packages, can be used to perform some application-level performance assessment, such as request response times. These tools attach to any software component or to the hypervisor and are able to observe all incoming and outgoing messages. What is common among all these approaches is that they generate the logging information on the end-hosts. The tools themselves can then either additionally perform the analysis, write the logging information to a file to be analyzed later, or send the logs to specialized analysis systems that run on a different set of hosts.  In any case, application performance may be affected by sharing the host resources with the logging and/or analysis tasks. 

In contrast to these traditional solutions, this paper explores how some of the application monitoring functionality can be moved to the network layer by taking advantage of
Software-Defined Networking (SDN) and Network Function Virtualization (NFV). SDN separates data and control flow, and NFV allows software implementations of network functionality. In fact, SDN already provides some support for targeted network monitoring such as bandwidth
and message statistics for individual flows.   SDN and NFV are invaluable for cloud data centers; according to Cisco's November 2018 cloud index report\cite{cisco} $1/3$ of cloud data center have partially or fully adopted SDN and $2/3$ will do so by 2021. By then SDN and NFV traffic volume will make up 50\% of the data traffic within the data center. Given such fast adoption, we believe it is important to explore how much SDN can be exploited as a low-level middleware to  support cloud-wide application monitoring. 

We have been inspired to do so by a number of recent studies that explore the feasibility of performing a  more global network monitoring approach based on SDN \cite{selectivemirror,SDN-PANDA,DAAS}. The most basic approach is to enable the port mirroring feature in switches and routers, in such a way that they mirror application traffic to specialized analysis systems. That is, messages are duplicated at the network component, and the actual analysis takes place somewhere else. Although such an approach is simple, and does not impose major computational overhead at the switch/router, the message traffic overhead can be significantly high. Such overhead can be reduced by filtering messages at the SDN switches/routers using forwarding rules that ensure that only packets that are important for monitoring are mirrored to the specialized analysis system. The disadvantage of selective mirroring is that it might lead to a delay in the forwarding path. 

Other research \cite{DPI, UMON, Ringbuffer, ApplicationAware, UDA, UDA2} extends the source code of the SDN switch to have an inclusive analysis function. However, the results show significant performance degradation in terms of the modified SDN switch forwarding throughput. In this paper, we have a closer look at some of these approaches and see how they can be adjusted to work for application monitoring. In particular, we analyze how effective selective mirroring is able to truly filter only relevant messages. We also provide insight into the performance  implications.  

Furthermore, driven by the deployment limitations of existing approaches, we propose a conceptually different solution that can be used for software switches  which run on general purpose hardware. Such software switches are often deployed at rack-level and/or for the VMs/containers hosted on the same machine.  In our solution,  we decouple the monitoring functionality from the switch forwarding path by developing a monitoring middleware component that runs on the switch node, yet it is only loosely coupled with the switch functionality. We show in this paper how such a loosely coupled approach offers a flexible and adjustable integration of application monitoring and analysis functionality into the switch. 
 In particular, we explore two variants of our middleware. The first does the required complete analysis locally on the switch host. The other performs some filtering and pre-processing of messages and then sends the necessary information to an analysis node, similar in concept to the selective mirroring approach described above. 
 



In summary, the contribution of this paper is as follows:
\begin{itemize}
\item We analyze various approaches in which network components are used in support of application monitoring, outlining their advantages and disadvantages. 
\item We propose a solution in which monitoring functionality is loosely integrated into software-based switch components, allowing for a flexible and adjustable deployment of application-based monitoring and analysis. 
\item We have implemented and evaluated our proposed solutions, and compared them against existing monitoring solutions that are deployed on the end-hosts on which the application is running. Our evaluation is based on the YCSB benchmark and analyzes the corresponding trade-offs in terms of impact on the application performance, general resource consumption, switching speed and communication overhead.
\end{itemize}

The remainder of the paper is as follows. Section~\ref{BG} provides background about SDNs, cloud network architectures and traditional end-host based application monitoring. Sections~\ref{collection}
and \ref{MA} present our network-based application monitoring solutions. Specifically, Section~\ref{collection} focuses on the role of the software switch and how relevant information can be filtered and transmitted to the monitoring component. Section~\ref{MA} then explains how application-relevant data can be inferred from the individual messages. Section~\ref{Evaluation} discusses the validation methodology and report the performance  results. Section~\ref{related} present related work and Section~\ref{Conclusion} concludes the paper.

\section{Background}\label{BG}

\subsection{Software Defined Networking}

In Software Defined Networking, the SDN controller represents the control plane that decides on how messages are routed and SDN switches represent the forwarding plane (data plane) that guide messages through the network. The controller provides each switch with a set of rules indicating how to forward the different flows of messages they receive from end hosts or other switches. A flow is typically identified by a set of IP header fields. Forwarding rules can change over time as the controller dynamically customizes how to route individual flows. OpenFlow -- the de facto standard of SDN -- is an API used for exchanging control messages between the controller and switches.  

Each SDN switch has one or multiple flow tables, configured by the SDN controller through the OpenFlow API. A flow table contains rules to match incoming flows and/or packets with certain actions such as prioritization, queuing, packet forwarding and dropping. 
The following is an example of a flow table rule, consisting of a set of conditions and actions to execute when the conditions are met: 
\begin{quote}
Conditions:
\\ \textit{TCP-protocol, Source-IP$=$A.A.A.A, Source-Port$=$X, Destination-IP$=$B.B.B.B, Destination-port$=$Y}  \\
Actions: \textit{output to out1}
\end{quote}
The rule will send the Acks of TCP messages sent from port $X$ of machine with IP A.A.A.A to port $Y$ of machine with IP B.B.B.B to out1, which is the switch's ports that is connected to the target destination. 

Network functions, such as packet switching but also more complex tasks, such as intrusion detection, load balancers and firewalls, are traditionally implemented as custom hardware appliances, where software is tightly coupled with specific proprietary hardware. Network function virtualization has been recently proposed to provide more agile and cheaper networks. This means that network functions -- such as packet switching -- can be implemented as an instance of a middleware software that is
decoupled from the underlying hardware, and running on standardized compute nodes. OpenVswitch (OVS) is one of the widely used virtualized software switches in today's cloud data centers. Various cloud computing platforms and virtualization management have been integrated with OVS, including OpenStack, openQRM, OpenNebula and oVirt.

\begin{figure}[t]
\centerline{\includegraphics[width=\columnwidth]{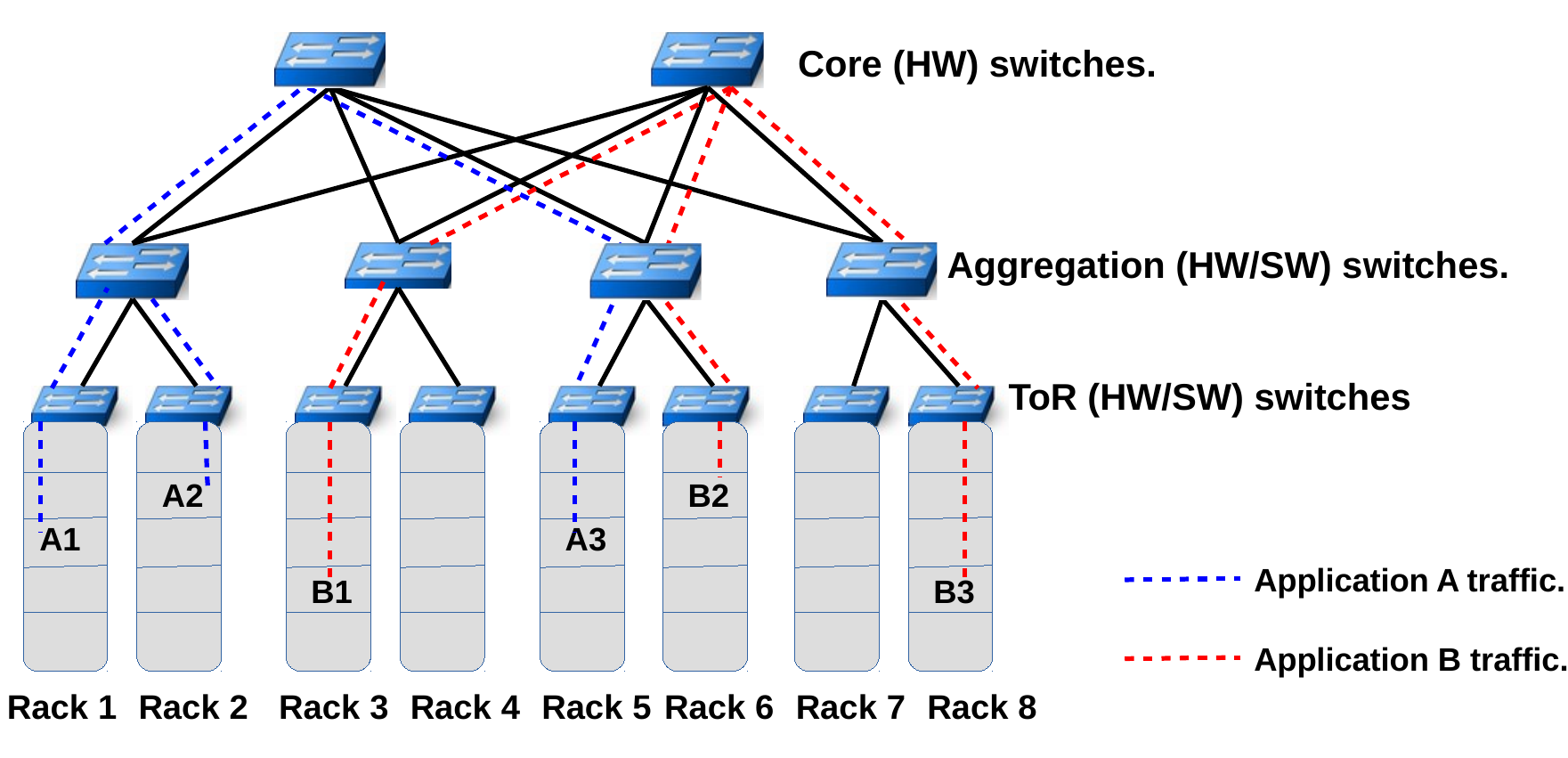} }
\caption{Articulation of a sample cloud network architecture}
\label{fig:CNA}
\end{figure}

\subsection{Cloud network architecture} A typical cloud data center network architecture today consists of a 2-3 layer tree of switches and/or routers (e.g., Fat Tree \cite{fattree}), such as shown in Figure~\ref{fig:CNA}. Hardware packet switches are being used in the core network  where low latency is a must. In contrast,  optimized software switches (e.g., OVS integrated with DPDK\cite{DPDK}, that accelerates packet processing workloads running on general-purpose CPU), are largely deployed as top-of-rack (TOR) switches ~\cite{netvm,DPDK2,netmap,clickos}. OVS is also used within high-end server machines that host many virtual machines and/or containers where the OVS instance routes the messages exchanged between the VMs/containers running within this same machine.


\subsection{Traditional Application monitoring}

\begin{figure}[t]
\centerline{\includegraphics[width=\columnwidth]{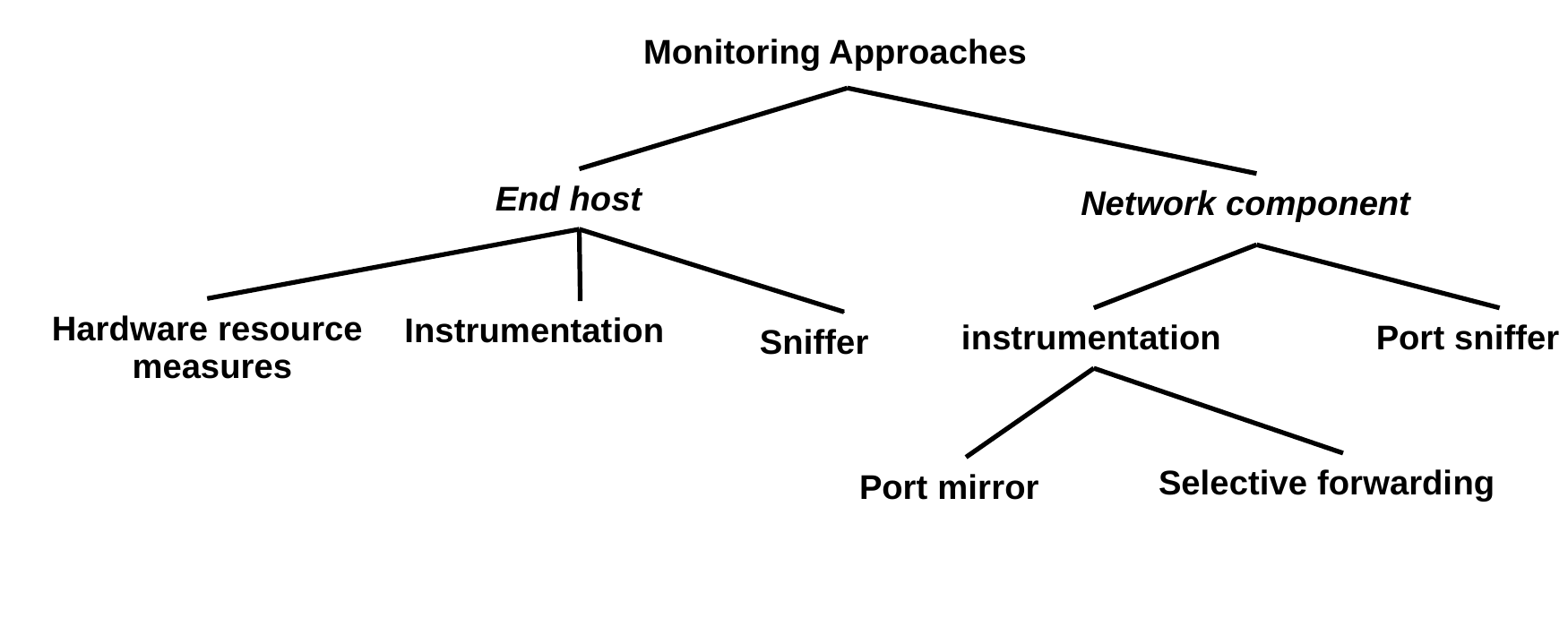}}  
\caption{Categorization of application monitoring approaches}
\label{Fig:ClassificationTree}
\end{figure}

 A distributed cloud application might be be spread across many nodes as illustrated in Figure~\ref{fig:CNA}, where the client request flows from one component to the next, each performing a different task to complete the request. Monitoring such an application requires both the \emph{collection} of relevant data  and the \emph{analysis} of the data to provide useful performance statistics. The left part of Figure~\ref{Fig:ClassificationTree} categorizes most of the existing solutions as they  extract the information relevant for monitoring at the end hosts.

 Assessing the utilization of hardware resource for the individual components in terms of CPU, memory and I/O is typically a first and straightforward step when analyzing performance \cite{systemlayermeasures}, and obviously,  all related measures need to be taken at the end hosts, where the components reside. 
 However, performance issues are not always caused by hardware bottlenecks, and there exist a wide range of high-level measures that are of interest, such as request service times for
individual components. This paper focuses on these higher-level performance metrics. 
 
 One way to extract such high-level metrics is through 
\emph{software instrumentation} which typically creates explicit application and/or platform specific log entries that then allow to extract the high level measures such as request service time \cite{SAAD,twitter,twitter2,semanticlogging}. 
For instance, Apache Tomcat uses its proprietary ``Access Log Valve" to create log files. Any request arriving at a Tomcat web application is passed to the access log valve process as well, and the same applies for responses. This allows the valve to calculate high-level measures such as response times. 
Software instrumentation is an application and/or platform dependent logging process. Obviously, Tomcat valves can't be used in a different servlet/JSP container. Thus, software instrumentation  will not support well monitoring of complex multi-tier applications that execute across various kinds of components. 

Network monitoring tools can be used to observe and monitor the message exchange between components. We refer to them as the \emph{sniffers} in Figure~\ref{Fig:ClassificationTree}.  Tools like Wireshark/Tshark or tcpdump are deployed at the end-host and can capture relevant messages in real-time and very efficiently by using message filtering at the interface to the application layer. They then sniff the relevant data of filtered messages, mainly the header information, and transform them into log entries. Most typically, these log entries are then written to files. Additionally Wireshark/Thsark provides quite sophisticated analysis tools for these messages, and some of the metrics provided are application relevant. One example is request service time. Wirehshark/Tshark can calculate it by matching an outgoing response message with the corresponding incoming request message it has seen before and taking the difference between the two capture times as the request service time.

Software instrumentation tools and network monitoring tools typically write their log entries to a file for offline analysis, perform the analysis locally in real-time and send the results for visualization (e.g., to a monitoring screen), or send their log entries to a remote analysis tool. 
 If  the analysis is done at the hosts, it might have a negative effect on the resources available for the application itself, and dedicated nodes will cause logging-related network traffic which might affect the performance on the end host and the network. 
 
 \section{Information Collection in the Network}
\label{collection}

\begin{figure*}[htbp] 
\centering
 \begin{subfigure}[c]{0.20\textwidth}
    \includegraphics[width=4cm]{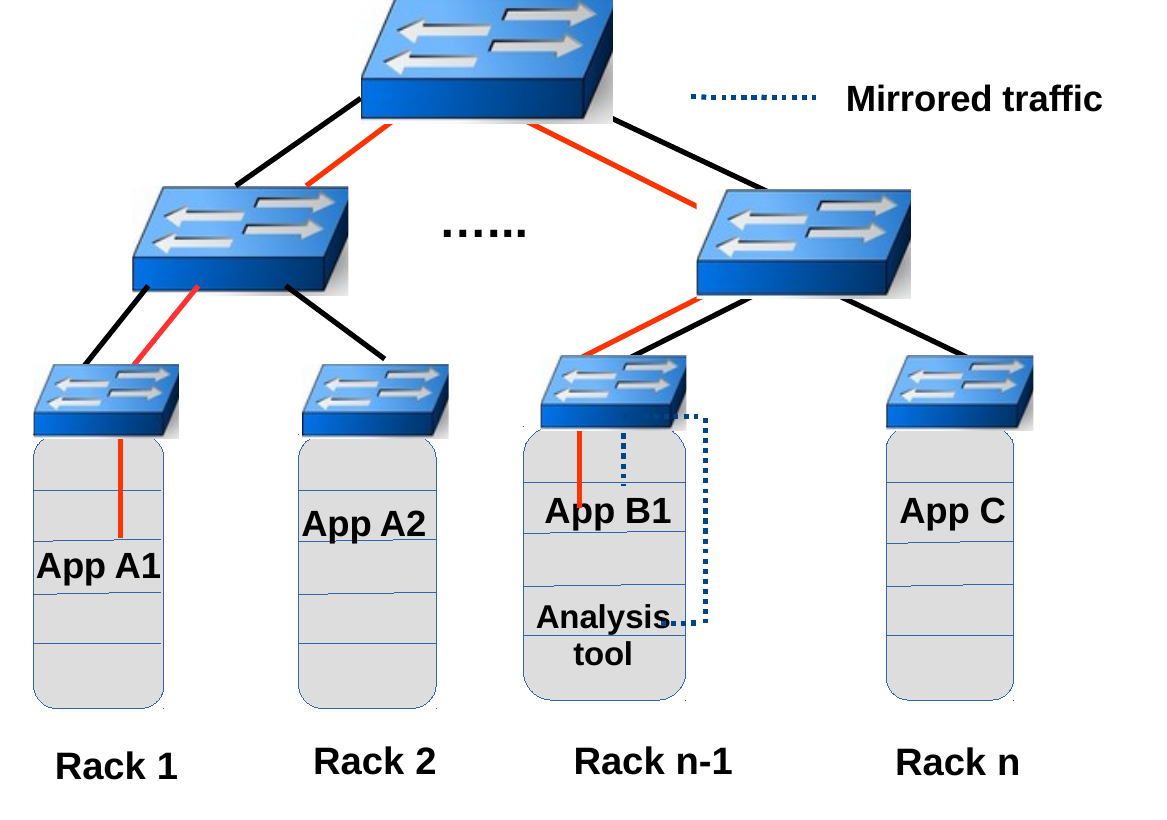}
     \caption{Port/Selective mirroring}
     \label{Fig:portmirror}
  \end{subfigure}
  \hfill
  \begin{subfigure}[c]{0.32\textwidth}
 \includegraphics[width=7cm]{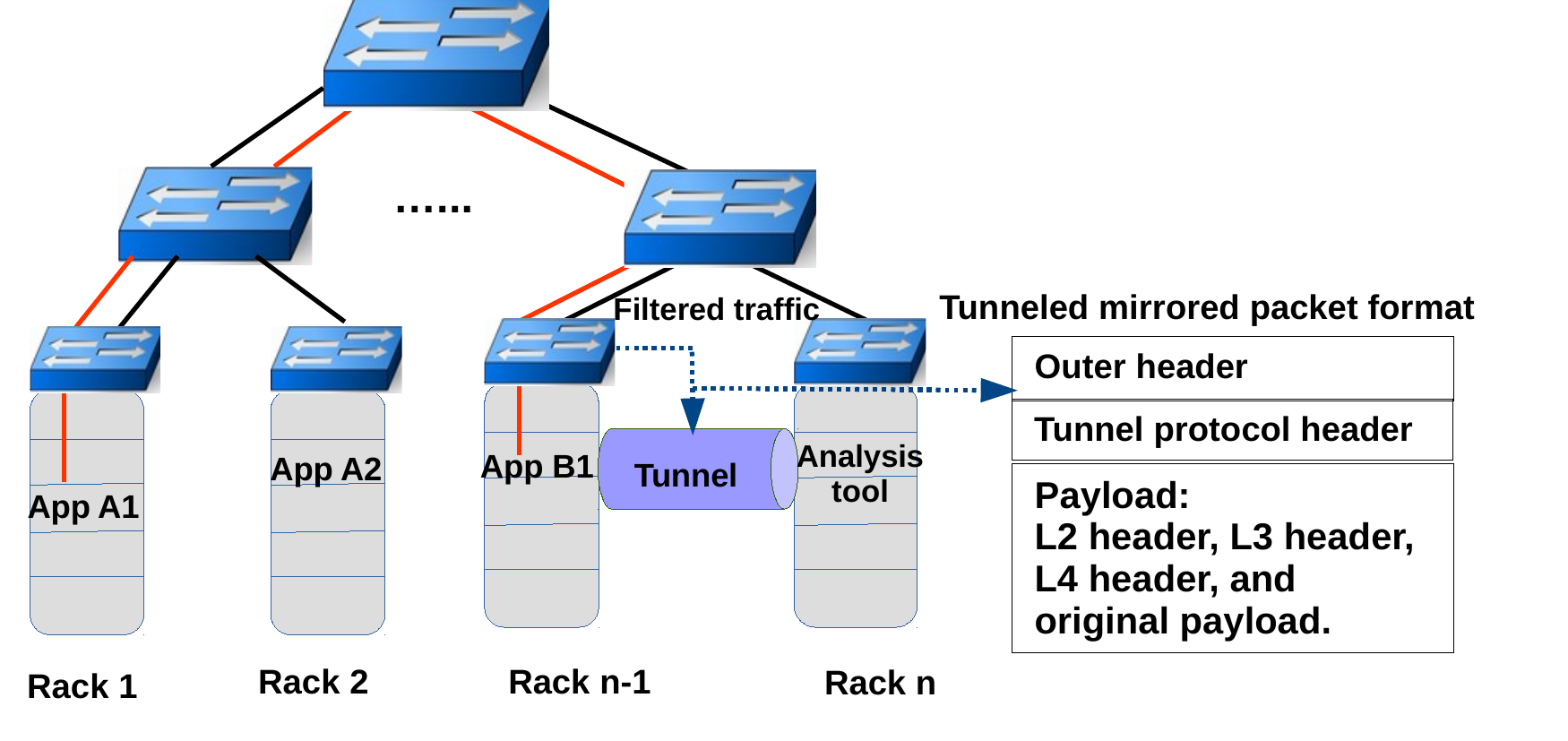} 
     \caption{Tunneled selective mirroring.}
     \label{Fig:selectivemirror}
  \end{subfigure}
  \hfill
  \begin{subfigure}[c]{0.32\textwidth}
\includegraphics[width=7cm]{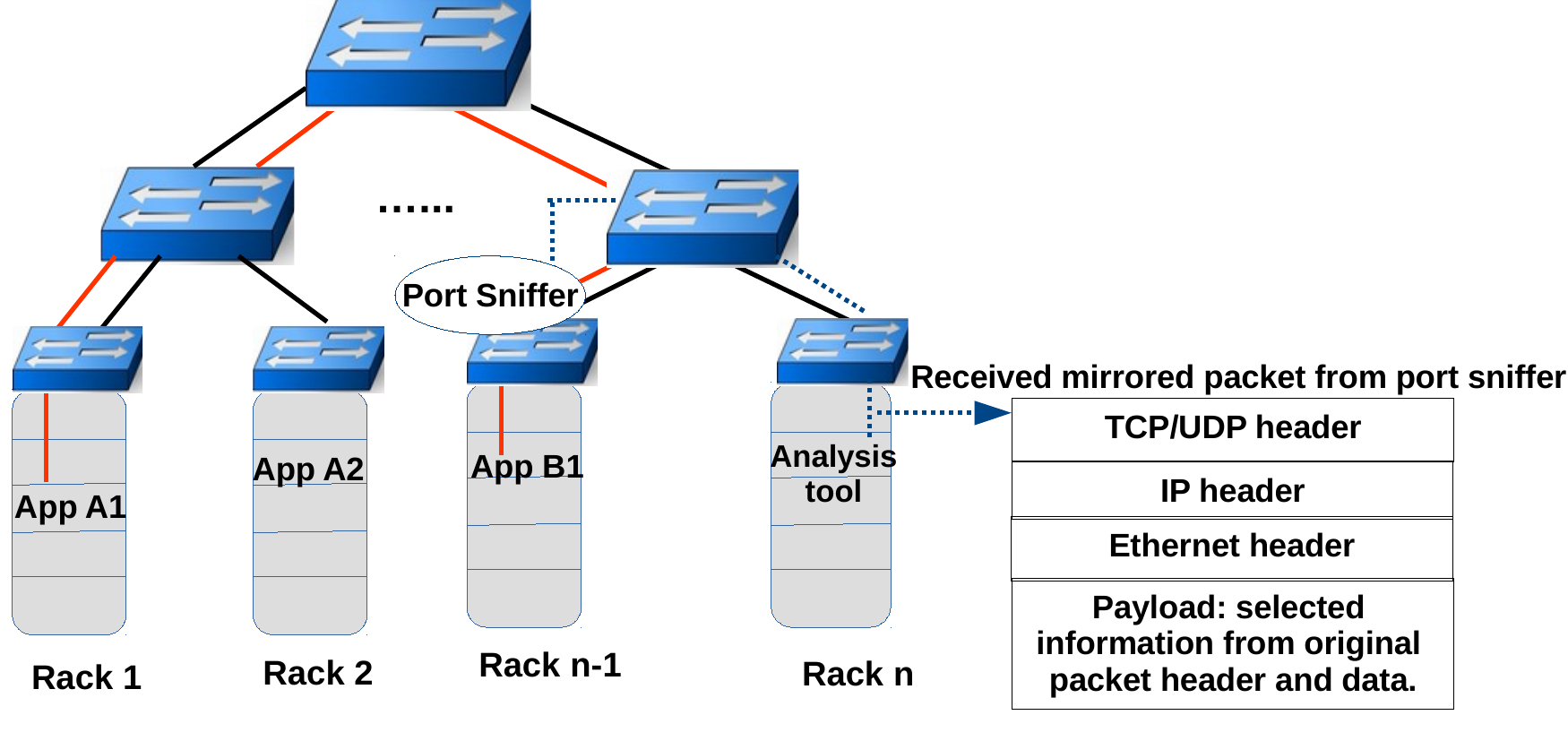} 
      \caption{Port sniffer.}
      \label{Fig:portsniffer}
  \end{subfigure}
\caption{Monitoring options in the Network}
\label{Fig:LoggingApproaches}
\end{figure*}

A key observation about the network monitoring tools mentioned above is that they provide some application specific measures by only looking at messages exchanged between components. Thus, the work performed by these tools can potentially also be done in the network. As messages travel through the SDN switches,  we argue that the switches can potentially be programmed to perform work relevant to log collection and analysis. In this and the next section we explore various alternatives to do so in detail.  They are depicted in the right side of Figure~\ref{Fig:ClassificationTree}.

Port and selective mirroring are approaches that have already been explored for network monitoring. What we provide in this paper is a detailed analysis of their capability to support application monitoring. 
The port sniffer is a new approach that we propose in this paper. With this approach the switch component has the flexibility to either send information to an analysis tool or do it itself. 

In the remainder of this section we provide an overview of the principles of each of these approaches and describe what exactly is done within the switch. The following section then focuses on the analysis itself; that is, which parts of messages are used to perform certain analysis and how the measures can be calculated.


\subsection{Selective port mirroring}
Using \emph{port} mirroring, the switch forwards all messages of a flow not only to the indicated destination but also to a secondary destination.   Figure~\ref{Fig:portmirror} shows an example of port mirroring of all network traffic between application component A1 and B1. As port mirroring runs with low priority, the performance of mirroring may be degraded or in extreme circumstances be temporarily suspended if there is a lot of normal packet traffic. Port mirroring is very fast as no copy process takes place but might have a significant communication overhead as it mirrors all packets of a flow. The application-performance monitoring framework developed in \cite{Timothy2015} uses port mirroring to forward a copy of all packets of a flow of interest to a monitoring agent that then extracts relevant information from messages of interest, and performs the analysis.

\paragraph{Reducing message overhead}
The number of packets to be mirrored can be reduced by using \emph{selective mirroring}. We are not aware of any application monitoring tool that currently exploits selective mirroring.  Selective mirroring uses the  filtering capability of OpenFlow compatible switches to copy and forward only those network packets to the analysis tool that match a predefined criteria, thus reducing the number of packets to be transferred. Furthermore, messages can be truncated so that only the first $X$ bytes are mirrored, with $X$ being a parameter.  Selective mirroring can be defined in the switch through the \emph{flow rules}. 

For example, suppose that A1 in Figure \ref{Fig:portmirror} is a frontend server, and B1 is the web server and we only want to mirror data packages to the analysis tool as only those are relevant to measure request service times. In order to avoid mirroring many of the control messages that are sent from A1 to B1, we can add two OpenFlow rules to B1's TOR switch, one for each flow direction (A1 to B1 and B1 to A1). The one that captures the message flow from A1 to B1 would roughly look like:
\begin{quote}
Conditions: \\
\textit {TCP-protocol, \\ Source IP $=$ A1's IP, Source Port $=$ A1's port, \\ Dest. IP $=$ B1's IP, Dest. port $=$8080, \\ TCP-Flags $==$ ACK|PSH} \\
Actions: \textit{Forward to B1 and Analysis Tool.}      
\end{quote}

In this scenario, only messages where the ACK or the PUSH flags are set are mirrored. In many of the control messages, these flags are not set. Furthermore, if only the header of a message is needed for the analysis, we can also use additionally the truncation option to mirror only the packet header, and avoid sending possibly large and mostly irrelevant payloads to the analysis tool. 

One has to note that selective mirroring is coupled with the switch forwarding path as the OpenFlow rules are executed at the time messages are processed to be sent to the actual destination.  This  may  lead  to  a  processing  overhead  and thus, delay in routing the message. 

Furthermore, the OpenFlow rules only apply to the header of the packets. A deep inspection is not possible. Thus, in some cases one might have to send more packets than actually necessary as a fine-grained filtering might not be possible. We will discuss in the next section some of the options and limitations of OpenFlow rules that would be relevant for application monitoring.

A major disadvantage of mirroring  and selective mirroring is that the machine hosting the analysis tool has to be directly connected to the switch. 
Mirroring does not change the message to be mirrored, i.e., it keeps its original destination IP and port in its header. Thus, if the analysis tool is not directly connected to the switch that performs mirroring but the mirrored message is instead sent first to a further switch, then this switch would actually forward the mirrored message to the  original destination; such a scenario is not desirable.  
Furthermore, the machine that hosts the analysis tool has to be put into a special mode as it has to be ready to receive messages that are not addressed to it. 

In the example in Figure~\ref{Fig:portmirror}, the switch to perform the mirroring is the rack switch on which the analysis tool runs; thus, there is a direct connection. Furthermore, in case of software switches that connect components running actually on the same physical machine, mirroring is also a possibility.


\paragraph*{Tunneled selective mirroring}
If the analysis tool should reside anywhere in the network, packet tunneling can be used. Tunneling allows private network communications to be sent across a public network, such as the Internet, through a process called encapsulation. The encapsulation process allows for network packets to appear as they are part of the public network, allowing them to pass normally. Tunneling has been used in the past for network monitoring such as Everflow~\cite{selectivemirror}, yet we are not aware of research that would exploit it for application monitoring. 

Tunnels, in conjunction with OpenFlow, can be used to create a virtual overlay network with its own addressing scheme and topology \cite{selectivemirror}. Figure \ref{Fig:selectivemirror} shows an example where the switch of B1 is programmed to tunnel messages to the analysis tool. Tunneling protocols such as GRE \cite{GRE} or VXLAN \cite{vxlan} encapsulate network data and protocol information in other network packet payload. An outer header is added to allow the encapsulated packets to arrive at their proper destination. At the final destination, de-capsulation occurs and the original packet data is extracted. Figure \ref{Fig:selectivemirror} illustrates this process, where the outer header contains the Ethernet and IP headers of the sending switch and tunnel destination, and the payload
contains the original packet (starting from the L2 header). As such, the mirrored packets can be sent anywhere in the network and different packets can be sent to different destinations.

To enable tunneling, the switch will be configured through the OpenFlow rules to set up a tunnel between itself and the host where the analysis tool resides. Then, OpenFlow rules similar to the one described earlier in this section are added to the switch in order to define the selection criteria and  encapsulate the network packets to send them through the tunnel to the analysis tool. This happens for request messages sent from A1 to B1 and for response messages from B1 to A1. At the endpoint of the tunnel, i.e., the analysis tool, first a de-capsulation has to take place before analysis can be started on the original data packets.

\subsection{Customizable Port Sniffer}


While mirroring and selective mirroring are only capable of forwarding (hopefully efficiently) relevant messages to an analysis tool that then does the actual analysis, we propose in this section a new approach that provides more flexibility  and allows the switch to perform some analysis locally. This requires the switch to be a software switch, e.g. the one based on OVS. We refer to this approach as \emph{port sniffer}. Software switches receive and send messages through ports. Thus, a sniffer process can be deployed at the ports to inspect all incoming and outgoing messages. This mechanism is depicted in Figure ~\ref{Fig:portsniffer}. For instance, assuming that all ToRs in Figure \ref{fig:CNA} are virtualized on computing nodes, a sniffer process can be deployed on B1's ToR switch host and instructed to capture all traffic traversing the virtual switch port connected to B1. This has conceptually similarity to what network monitoring tools such as Wireshark/Tshark do. 

The sniffer is an independent process on the node running the software switch. In principle, the sniffer can implement any kind of semantics; for example, the sniffer may simply forward selective messages to an analysis tool, aggregate and reformat logging messages that only contain information relevant for the analysis, or just perform the analysis by itself. In principle, one could even deploy tools such as Wireshark on the switch node. 
In our implementation, we follow a flexible approach that allows a wide range of possibilities. 

Our port sniffer separates the actual capturing of messages from any additional tasks. The \emph{listener process} keeps sniffing on predefined switch ports, filters relevant messages, and saves the needed traffic packets into a shared memory space. From there, further extraction, analysis and forwarding are performed by extra process(es) as needed. We have implemented the listener in a separate process as it has to work at the speed of the OVS ports. Thus, we wanted to make the listener task as simple as possible, allow for straightforward multi-threading and avoid interference with analysis functions.  

For example, to measure HTTP service request times, the listener process sniffs the OVS port that is connected to the web server and filters the relevant traffic between client and server (i.e., network packets with port 80 or 8080) to be analyzed.
From there, we have implemented two versions for further processing. One performs the calculation of request response time locally within an \emph{analysis process}. In the second version, the \emph{extract and forward process} of the sniffer extracts again the relevant information, determines timestamps and data packets, but does not do the matching itself. Instead this time the information is forwarded to a remote analysis tool, similar in concept to what the selective mirroring is doing. It uses UDP for that purpose.

\section{Message Analysis}\label{MA}
\begin{figure}[t]
\centerline{
\includegraphics[width=\columnwidth]{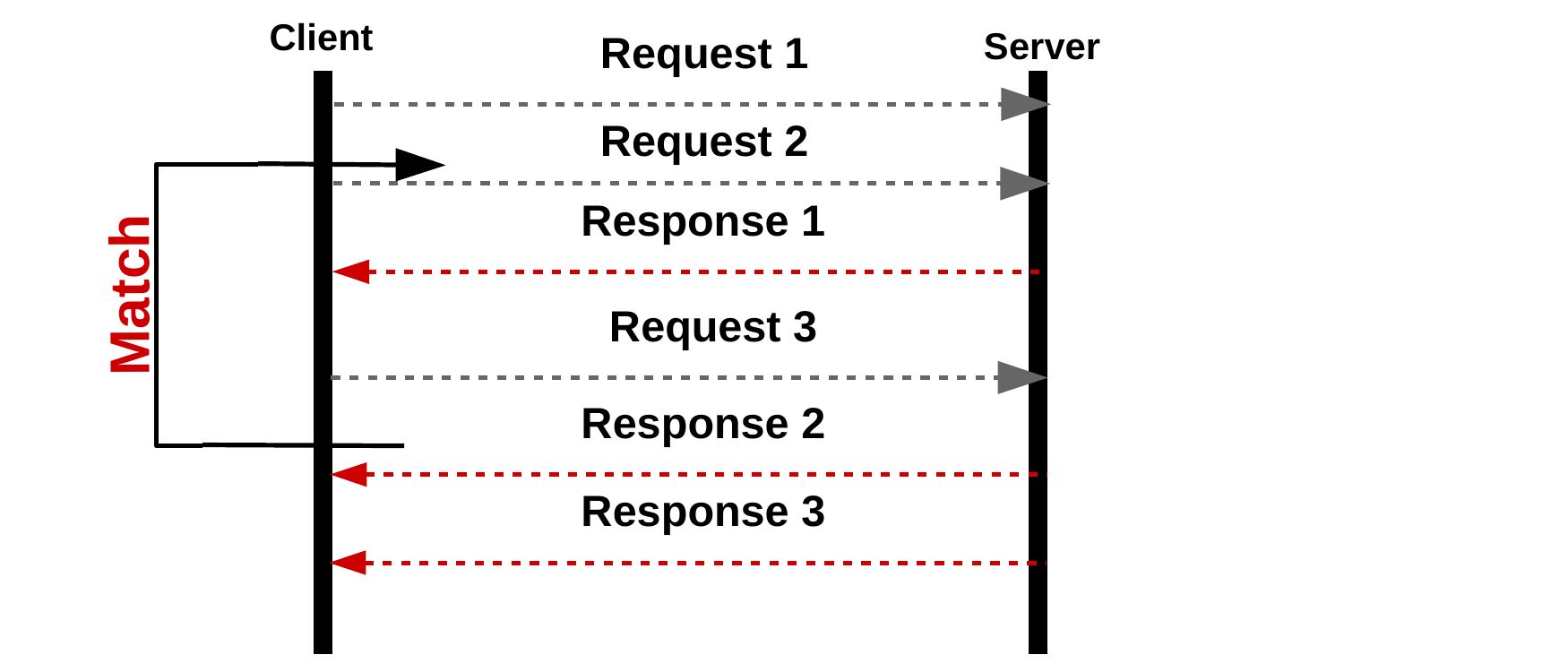}  }
\caption{ Request/response packet pair matching methodology}
\label{Fig:HTTP}
\end{figure}

As pointed out above, considerable application relevant measures can be extracted by looking at message content providing information about the performance of the individual components and the system overall. In this section we want to motivate the possibilities by outlying some of the measurements we can do through message inspection, and how we have implemented them for the different approaches that we presented above.
\paragraph{Request service time:}
In most component-based systems,  a component (or the client) uses a request/reply protocol to call the service of a different component. 
In many of these protocols, in particular http, a client connection to the server can have at most one outstanding request; that is, a client can only send a new request once it has received a response for the outstanding request. Thus, by having access to the flows from client to server and from server to client, one can take the time difference between the observed request and response as the request service time. If a client is allowed to have multiple outstanding requests (as shown in Figure \ref{Fig:HTTP}), then one can simply assume that the first response refers to the first request, the second response to the second request, etc. In fact, as also the client needs to know to which request to match a response, some servers guarantee that they will send responses only in the order they received requests even if they execute the requests concurrently.

Most request/reply protocols, including http, use TCP as underlying communication mechanism. Thus, the header information of messages follows TCP format, and the http (or other protocol specific) headers are within the data payload. 
Requests and responses could be spread across several TCP messages. While unlikely, even the protocol specific header could be spread across more than one TCP message. Thus, we have to be careful of how we get the right messages.

\begin{itemize}
    \item \emph{Using deep inspection:}
Deep inspection occurs when we look at the message payload past the TCP header. 
As TCP headers have a fixed size, it's quite straightforward to perform deep inspection and extract the protocol specific headers. Our port sniffer approach performs such a deep inspection. It detects the TCP packet that contains the last part of the http request header for the flow from client to server, and the packet that contains the last part of the http response header for the flow from server to client. If analysis is done within the sniffer, it then measures the time between the arrival of these two packets. Otherwise, it simply forwards this information to the analysis tool which then takes the time difference between receiving these two messages. 
\item \emph{Using TCP headers:}
Selective mirroring cannot do deep inspection but has only access to the TCP headers.
Therefore, in our selective mirroring approach, our filtering rules rely on the assumption that if a request or reply message is split into  $n$ TCP-packets ($n \geq 1$), then in the last of these TCP packets, the TCP header flag "PUSH" is set to true. We have confirmed that this assumption holds for the http implementation we have deployed. With this, our rules filter for each flow direction only packets with
\begin{quote}
    TCP-Flags $==$ PSH
\end{quote}

As a result, for both the request and the reply message, exactly one packet is mirrored to the analysis tool, and the analysis tool takes the time difference between receiving these two messages as the service request time.

\end{itemize}

As we might want to observe many different connections at the same time using selective mirroring, we have to typically define flow rules for each connection. However, sometimes, it is possible to define wildcards, e.g., to mirror all messages that are sent to a server independently of the sender.  

In the analysis component, whether it resides within our customizable port sniffer or is on a separate node,  we have to distinguish the messages from the different connections of interest. Thus, when information about a request message arrives it is stored in a connection specific data structure until the response arrives. Only the time difference between a request and its response needs to be kept track of. If several requests are queued when a response arrives, the match is done with the oldest request as depicted in Figure~\ref{Fig:HTTP}. 

Note that the precision of the measurements might depend on where timestamps are taken. For the mirroring approaches,  the times are taken when the mirrored messages arrive at the analysis tool. For port sniffing, the sniffer process can take the time. In both cases, this is not the time when the message was sent by the original source nor the time the destination receives the message. For example, using port sniffing, the time taken by the sniffer for the request is before the message arrives at the server, and  for the response it is after the message is sent by the server. Our assumption is that message delay times in the network are negligible compared to request execution times, especially if the switch in charge of mirroring or sniffing and the analysis tool are close to the server under observation. 

\paragraph{Aggregated request service times:}
Typically, administrators are interested in aggregated information. It is quite easy to calculate average, maximum and minimum service time over an observation window. Or in case of long-lasting observations, values can be given periodically as aggregates over predefined observation windows. Space and computation overhead to keep track of such aggregated information for each connection under observation is very small; thus, it should be possible to maintain them even at high throughput rates. 

\paragraph*{Server Load:}
To measure the load of a specific server, the filtering rules at the switch have to consider either all flows where the server is the destination and then count the number of different requests per time interval, or consider all flows where the server is the source, and then count the number of different responses per time interval. 
When using Openflow rules, we can use wildcards for the client IP to cover more than one client.

\paragraph*{Success Rate:}
Request/reply protocols typically have a status code for  each response message that identifies whether it has been successfully processed or not. By comparing successfully executed requests with all submitted requests, we can easily calculate success rates. However, the status code is part of the protocol header which is contained in the TCP packet payload and not in the TCP header.  Therefore, deep packet inspection is required to extract the response status code.

As such, for our selective mirroring approaches, as we cannot do deep inspection, we have to make sure that we sent all data packets to the analysis tool. Therefore, only sending TCP packets where the PUSH flag is set is no more sufficient, we need to send all data packets. Unfortunately, there is no specific flag in the TCP header that would indicate that a packet is a data packet and OpenFlow also does not allow to specify a rule that would allow us to filter only packets that have a non-empty payload. Therefore, we filter all packets that either have the ACK or the PUSH flag set. In our observations, all data packets have the ACK flag set (thus, we should not miss any of them), and most ACKs are piggypacked on data packets (therefore, we should hopefully not send too many unnecessary messages.

\paragraph*{Frequency of requests for individual objects/methods:}
As HTTP request messages contain the full URL, we can easily keep track of the number of requests per URL, calculating frequencies on an interval basis. For this, we simply have to parse the first line of the HTTP header. Again, this requires deep inspection of the packets.
    
\section{Evaluation}\label{Evaluation}
 \begin{figure}[t]
\centerline{
\includegraphics[width=\columnwidth]{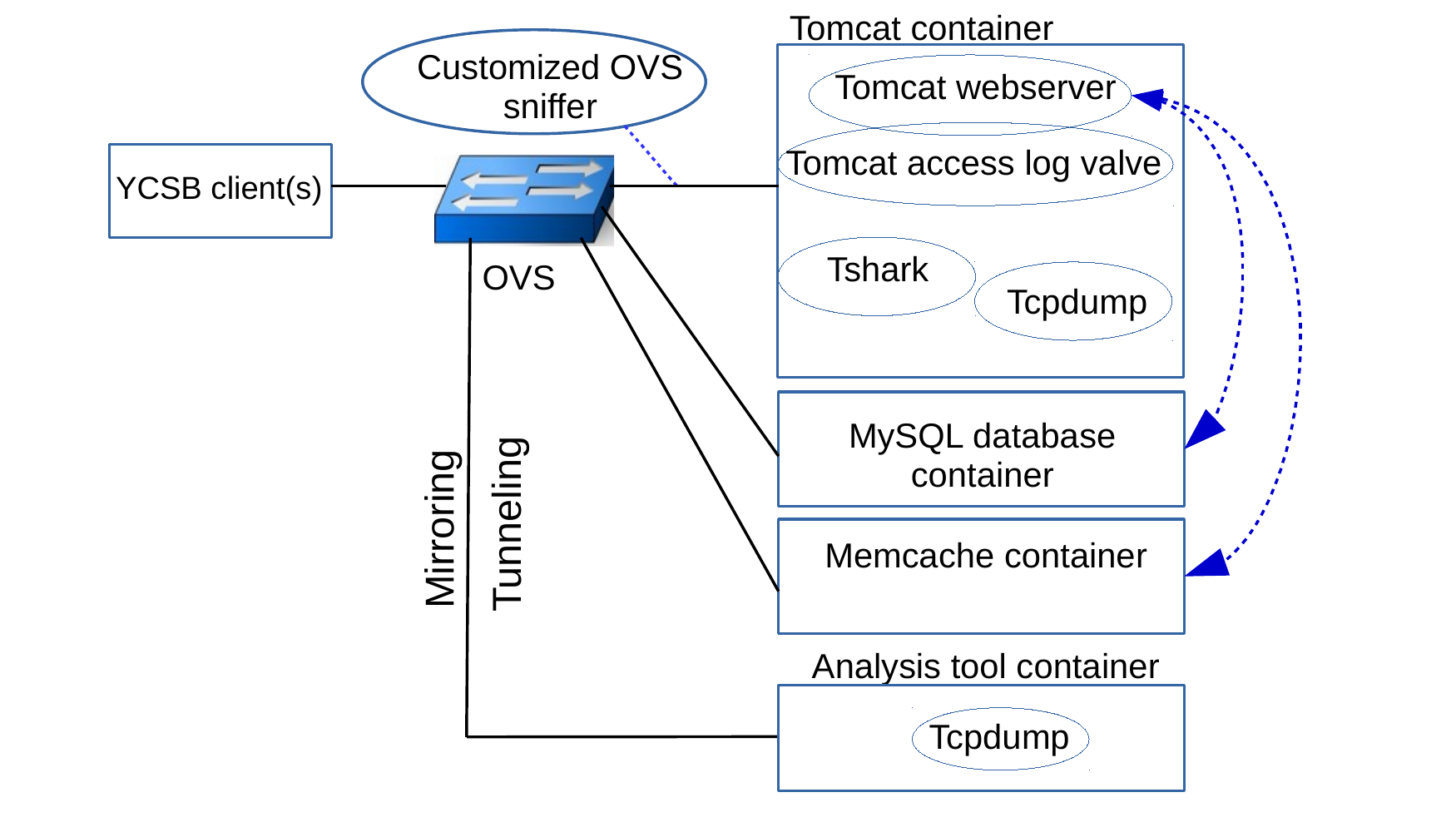} }
\caption{Test application architecture}
\label{fig:TestApp}
\end{figure}

\begin{table*}[t]
\caption{A list of evaluated Monitoring Approaches along with their characteristics} 
\resizebox{\textwidth}{!}{%
\begin{tabular}{*{5}{c}} 
\textbf{Monitoring approach}     & \textbf{Application dependent}       & \textbf{Deployment location}    & \textbf{Onsite analysis?}   & \textbf{Output}   \\ 
\hline  
\textbf{\textit{Tomcat logging}}  & Yes &       server host        & Yes &    Dump analysis results to end-host file \\
\hline
\textbf{\textit{Tshark}}  & No      & server host  & Yes &       Dump analysis results to end-host file \\ 
\hline 
\textbf{\textit{Tcpdump}}  & No      & server host  & No &      Dump captured data to end-host file \\
\hline 
\textbf{\textit{Customized OVS sniffer}}      & No  &  OVS host & Yes/No & Onsite: dump analysis result to OVS host file,\\
& & & & Offsite: dump captured data to analysis file   \\
\hline 
\textbf{\textit{Port mirroring }}     & No  & OVS host & No &  Dump captured data to analysis file \\ 
\hline 
\textbf{\textit{Selective mirroring }}     & No  & OVS host & No &  Dump captured data to analysis file \\
\hline 
\textbf{\textit{Truncated mirroring }}     & No  & OVS host & No &  Dump captured data to analysis file \\ 
\hline 
\textbf{\textit{Tunneling-GRE }}     & No  & OVS host & No &  Dump captured data to analysis file \\ 
\hline 
\textbf{\textit{Tunneling-VXLAN}}      & No  & OVS host & No & Dump captured data to analysis file \\ [1ex] 
\hline
\end{tabular}
}
\label{table:evalogtool} 
\end{table*}

In this section we present an evaluation of the approaches presented in the previous section and compare their performance also with some of the approaches deployed at the end hosts. 
Table \ref{table:evalogtool} provides characteristics of each evaluated monitoring approach  such as application/platform dependency, where information capturing takes place, whether the data collection and analysis are done at the same location, and what the output is provided by our implementation.

\subsection{Details of the Monitoring Approaches}\label{Eusecase}



For a platform-specific software instrumentation at the end host, we  enabled the access log valve in Apache Tomcat server to log HTTP request service times. We refer to this as \emph{Tomcat} in the performance graphs. 
For networking monitoring tools deployed at the web-server host, we use \emph{Tshark}, the command line interface to Wireshark, and \emph{tcpdump}. With Tshark, we can do the analysis in an online fashion, i.e., Tshark sniffs the messages, analyzes requests and either visualizes them or logs them to a file. Visualization was considerably more expensive. Thus, our evaluations show the overhead when results are dumped to a file.  Tcpdump is only a message capturing tool with filtering capability. Such capability is only used to filter messages that are relevant for the analysis. Tcpdump does not have an analysis engine, and is thus only instructed to dump all packets to/from the web-server port to a disk file, that can be fed into any offline analysis tool. 

For network-level application monitoring, we have evaluated port mirroring, selective port mirroring with and without truncation, tunneling using GRE/VXLAN and our customized OVS port sniffer. As both tunneling approaches have very similar performance results, we only show VXLAN in the graphs for better readability. For all approaches except of our OVS sniffer, the mirrored packets are sent to the analysis tool. For our customized OVS sniffer, we show the results when the OVS sniffer performs the analysis itself and when it sends relevant data to the analysis tool (similar to what the selective mirroring approaches do). The remote analysis tool used for mirroring and by our OVS sniffer actually only dumps all the messages it receives to a file without further analysis. We do so because in our setup the analysis tool resides on the same physical machine as the web server. To avoid indirect effect to the web-server we have tried to keep the overhead as small as possible. 


\subsection{Implementation Environment}

Our basis for evaluation has been the YCSB benchmark on an extended architecture as depicted in  Figure \ref{fig:TestApp}. While YCSB is originally a database benchmark where a YCSB client sends requests to a database, our extended version has added a Tomcat webserver as frontend for the client (which was modified to communicate with the webserver); the webserver has access to a MySQL database and a Memcache server. All components are connected by an OVS switch configured via Openflow. The clients submit a predefined workload of HTTP requests to the web server whereby each request retrieves data from either the database or the memory cache. Recent results are cached in the Memcache server. The database schema and the query requests follow the YCSB benchmark. A separate analysis tool component is used for some of the evaluated approaches.

The experiments are performed using DELL hosts with dual Intel(R) Xeon(R) CPU E3-1220 v5 @ 3.00GHz CPUs (4 cores per socket),a Broadcom NetXtreme BCM5720 Gigabit Ethernet Dual Port NIC, and 32.8GB memory, with the clients on one machine and all server components on another machine together with the OVS software switch. This resembles the scenario where the cloud provider has large end-host machines that host many components. 
Each server component runs in their own docker container  (docker-ce version 18.03.1) with predefined available resources and on separate cores.  All docker containers are connected by 10 Gigabit Ethernet OVS ports. We used OVS version 2.9.90. 16GB of RAM are assigned to the cache and the backend database system is MySQL 5.7.24.

In order to compare the performance of different monitoring approaches, we run our YCSB benchmark with and without monitoring. We then measure the overhead for each of the monitoring approaches by analyzing the client perceived performance. That is, we check how the different approaches affect the throughput and the latency observed at the clients. We also gauge CPU utilization for each monitoring approach and how  OVS forwarding performance is affected in case of the network-based approaches.

For our experiments, we chose the YCSB \textit{read only workload} with 3 million scan requests and zipfian distribution for records selection over a 10GB database (10 million records).   Each test scenario runs that workload for two minutes and the results are averaged for 5 runs. We tested with up to 30 client threads (after which the web-server was saturated even without monitoring enabled).

\subsection{Application throughput and latency}
\begin{figure*}
\begin{subfigure}{0.90\linewidth}
\centering
\includegraphics[height=3cm,width=\linewidth]{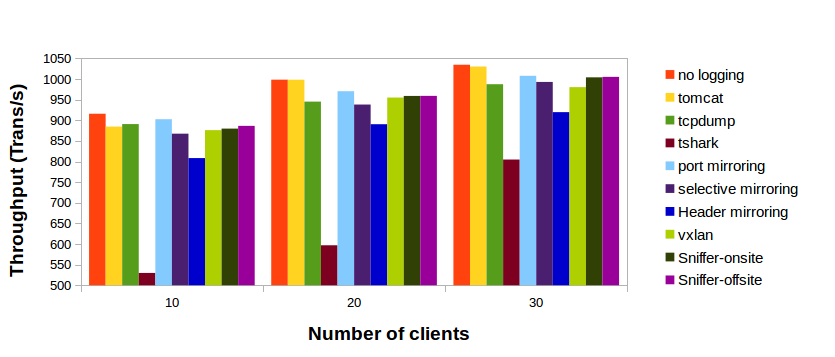}
\caption{Average throughput (transactions/second) reported by YCSB client}
\label{fig:Averagethrough}
\end{subfigure}\hfill
\begin{subfigure}{0.90\linewidth}
\centering
\includegraphics[height=3cm,width=\linewidth]{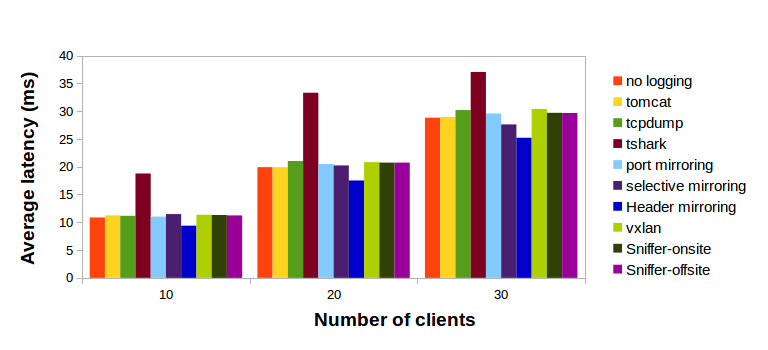}
\caption{Average latency reported by YCSB client}
\label{fig:AverageLaten}
\end{subfigure}
\caption{Application throughput and latency using various monitoring approaches}
\label{fig:throuandlaten}
\end{figure*}

Here we examine the impact of monitoring on the throughput and latency at the YCSB client. Ideally, monitoring has little to no impact on the performance observed at the client side.  Figure \ref{fig:throuandlaten} shows the end-to-end throughput and latency observed by the YCSB client with and without the integration of the various monitoring approaches and while increasing the workload, i.e.,  adding more client threads. 

Of all approaches, Tomcat works the best having nearly no negative impact on performance. This is because the access log valve does not need to perform any sophisticated message analysis; instead, likely through interception, it only records the time before the web-server starts processing a request and once it has completed, and then logs the time difference as service time. This is only possible because the valve is tightly integrated into Tomcat's software. The other two end-host mechanisms, both application independent, negatively affect performance. Tcpdump has much lower impact than Tshark, though. For example, with 20 clients tcpdump has 5\% lower throughput compared to no monitoring while Tshark has 40\% less throughput. This might be because tcpdump does not perform any analysis; it could also just be that Tshark has generally not an efficient implementation. 

All the network-based approaches perform significantly better than Tshark in terms of both throughput and response time (by at least 15\% for response time, and 30\% for throughput). In terms of throughput selective mirroring with truncating is at least 5\% worse than all other approaches (except of Tshark). Selective mirroring and tunneling are only insignificantly slower than tcpdump (by 1\%). In contrast, the throughput of port mirroring and our OVS sniffer solutions are as good as that of Tomcat. 

These are very promising results. Note that tcpdump stores the logged messages locally and the file needs to be retrieved from there before analysis takes place. In contrast, for the network-based approaches, we send all the relevant information to the remote analysis tool for online analysis or even perform the analysis on the fly such as Tshark.

To understand the performance differences and have a full picture, we look at resource consumption and the delay induced by mirroring at the OVS as well as some of the implementation details in the next sections. 
\subsection{Computational Overhead}\label{resourceconsum}

\begin{figure*}
\begin{subfigure}{0.49\linewidth}
\centering
\includegraphics[height=3cm,width=\linewidth]{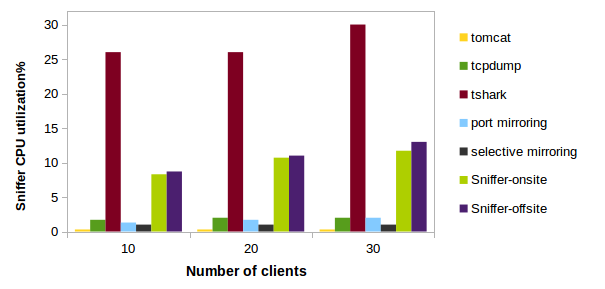}
\caption{Analysis tool CPU utilization}
\label{fig:ToolCPU}
\end{subfigure}\hfill
\begin{subfigure}{0.49\linewidth}
\centering
\includegraphics[height=3cm,width=\linewidth]{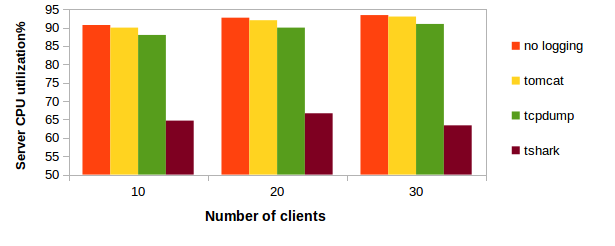}
\caption{Web server CPU utilization}
\label{fig:webservercpu}
\end{subfigure}
\caption{CPU overhead for different monitoring approaches}
\label{fig:CPU}
\end{figure*}

Figure~\ref{fig:ToolCPU} shows the CPU overhead of the analysis tool process, again with increasing number of clients. With Tomcat and Tshark, the analysis runs in the same docker as the webserver. With OVS sniffer, the analysis tool resides within the OVS host, and  for the rest of the network-based approaches, the analysis tool resides in its own docker. We used Linux top command to measure the CPU utilization of each running process inside the webserver and analysis tool dockers.

Figure~\ref{fig:ToolCPU} shows that Tshark has the highest CPU utilization and runs at full capacity. As the webserver docker is assigned to only one core in our experiments, Tshark shares such a core with the webserver process. Tshark consumes on average about 27\% of the CPU resources of the  webserver docker. This is the reason for the poor client-perceived performance. Note that Tshark has serious performance problems and frequently crashed during experiments. In addition, it also missed messages at higher rates. 

Our OVS onsite sniffer performs the analysis locally and writes it to a local file; it consumes 10\% of the OVS host CPU resources, which is much more efficient than Tshark. This may be because it does overall less analysis than Tshark. We note, however, that in our evaluation tests, we enabled only the Tshark analysis features that correspond to the ones performed in our OVS sniffer. Nonetheless, the OVS sniffer requires more CPU time than approaches that only dump data such as tcpdump which is very much expected. Interestingly, when the OVS sniffer only reformats messages and sends them to a remote analysis tool, it requires on average about 6\% more CPU utilization compared to the onsite sniffer version. It seems like creating messages and sending them to a remote site is more CPU intensive than performing the analysis locally.  

Generally, the CPU overhead is very low when messages are simply dumped to a file, about 1\% or less. The overhead is lower with selective mirroring than with port mirroring or for tcpdump, because the analysis tool only receives a subset of all messages with selective mirroring . This will be further validated by the communication overhead measurement in upcoming subsection.

Figure \ref{fig:webservercpu} shows the effect of sharing the webserver resources with the monitoring tool in traditional application monitoring approaches. The CPU percentage taken for the monitoring tool is deducted from the webserver process CPU resources, which diminishes throughput as shown in the previous subsections. The webserver CPU resources are decreased by 0.7\%, 3\% and 30\% for Tomcat, tcpdump, and Tshark;  respectively. 

\subsection{Switch overhead}\label{switchthrou}
In this subsection we compare the performance of port mirroring, selective mirroring, truncated mirroring and tunneling in terms of their impact on the OVS forwarding performance. To do that, we use Iperf \cite{Iperf} to measure core link performance. We deploy the Iperf server, the Iperf client and an analysis tool process, each in a separate docker container, all connected through 10 Gigabit Ethernet OVS ports. We work within a single  host as the focus is on the performance of the OVS software. We run experiments with up to 10 concurrent client connections. 

\begin{figure}[t]
\centerline{
\includegraphics[width=\columnwidth]{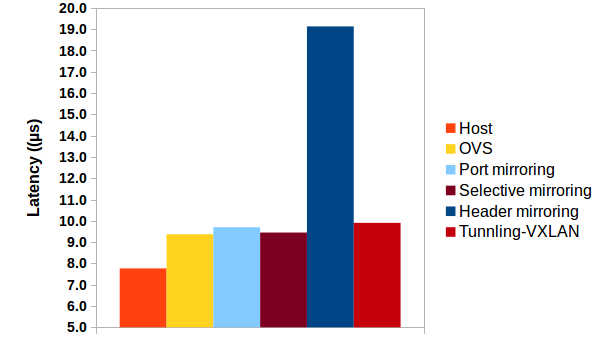}}
\caption{OVS link latency impact of different network-based monitoring approaches}
\label{fig:ovslatency}
\end{figure}

Figure ~\ref{fig:ovslatency} shows  forwarding latency when 10 clients are directly connected with the server (no OVS),  using OVS without any mirroring or tunneling, using OVS with port, selective and truncated mirroring, and using OVS with tunneling (VXLAN) to the analysis tool. As expected, direct connections without OVS perform best. OVS adds around 1.5 microseconds latency. Port mirroring, selective mirroring and tunneling are only slightly worse than OVS.  Truncated mirroring is the slowest.

With this, we can see that adding rules to OVS and even reformatting messages, as needed by tunneling, only add very little overhead to OVS.
Therefore,  the client-perceived performance impact that we discussed in Figure~\ref{fig:throuandlaten} is so small. 

However, truncating messages has a serious impact on the switch latency. In fact, not shown in a figure, the OVS CPU utilization is also 50 times higher for truncated mirroring compared to the other mirroring approaches.


\subsection{Impact of Implementation}

\begin{figure}[t]
\centerline{
\includegraphics[width=\columnwidth]{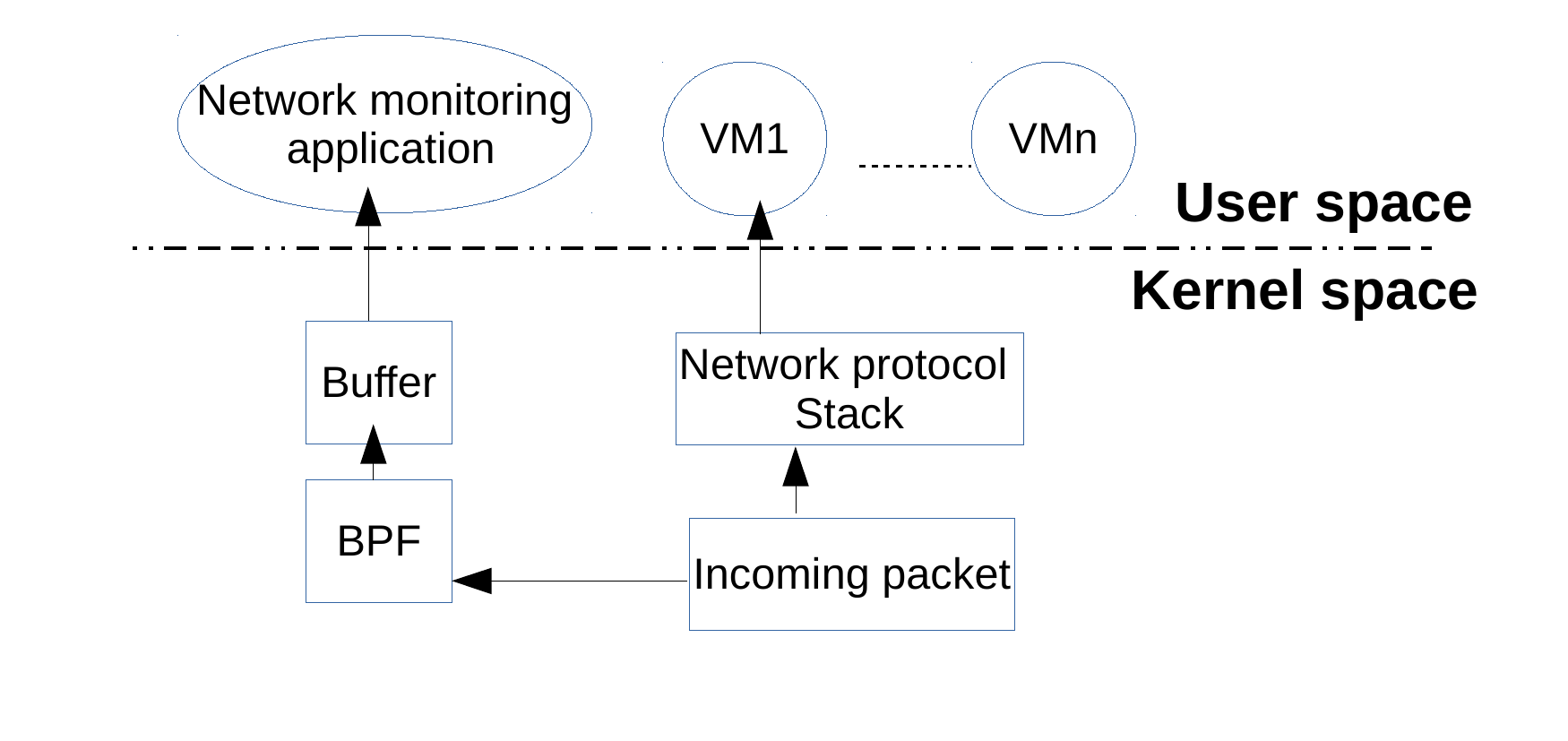}} 
\caption{Packet capturing overview}
\label{fig:kerneldelay}
\end{figure}

In general, for all network-based logging approaches, there is a  delay induced by copying messages, but in different contexts. As we mentioned before, in port mirroring, selective mirroring and tunnelling, the switch itself makes a copy of the captured packets to another defined port. In our sniffer, the kernel makes the packet copy for the monitoring application. The  packet capturing is done by inserting some filtering code (such as BPF \cite{BPF}) into the kernel at run time; this code will copy each incoming packet at the monitored port and send it to a buffer where the userspace monitoring middleware (such as our customized sniffer, tcpdump, tshark, etc.) will read the buffer and get the packets. Figure \ref{fig:kerneldelay} illustrates this process. While testing the customized sniffer, we have noticed an increase of kernel processing time. Thus, although with our port sniffer, the copying process is not in the OVS forwarding path to the original destination, we believe that such extra copying time in the kernel  has an indirect impact on the delivery rate of the packets, and thus, also on the client-perceived performance. 

\subsection{Communication overhead}
\begin{figure*}
\begin{subfigure}{0.49\linewidth}
\centering
\includegraphics[height=3cm,width=\linewidth]{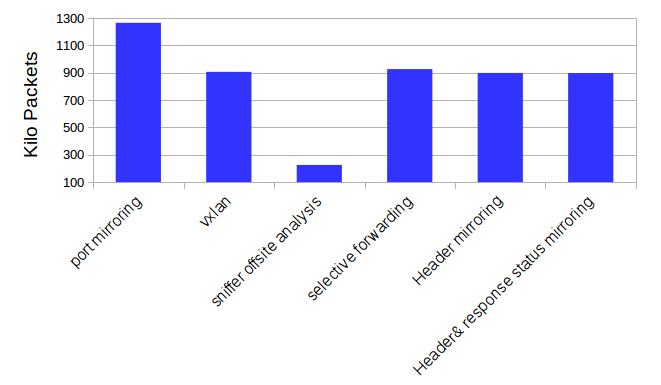}
\caption{Number of packets received by analysis tool}
\label{fig:sentpackets}
\end{subfigure}\hfill
\begin{subfigure}{0.49\linewidth}
\centering
\includegraphics[height=3cm,width=\linewidth]{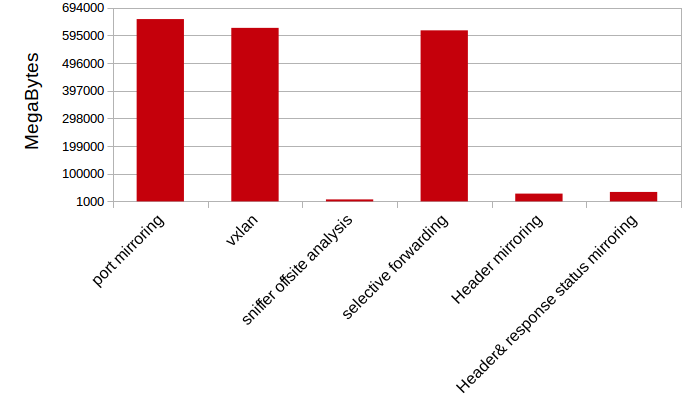}
\caption{Number of MBytes received by analysis tool}
\label{fig:sentbytes}
\end{subfigure}
\caption{Communication overhead when the analysis is conducted by a remote tool}
\label{fig:commoverhead}
\end{figure*}

In this subsection, we analyze the communication overhead induced by mirroring messages to a remote analysis tool.  We have collected the number of received packets and bytes at the analysis node while running YCSB with 20 clients. Figures~\ref{fig:sentpackets} and \ref{fig:sentbytes} show the number of transmitted packets (in 1000) and the total number of MBytes transmitted;  respectively.

Figure~\ref{fig:sentpackets} shows that port mirroring transmits the most packets to the analysis tool as it mirrors all packets over the monitored link. The OVS offsite analysis sniffer transmits the least number of  packets to the analysis node as it sends exactly one packet for each http request and response message. With this, it sends less than 12\% of the number of packages sent with port mirroring.  The selective mirroring variants transmit all similar number of packets because they use the same filtering switch Openflow rules. Compared to port mirroring, they have around 30\% less messages.

Figure\ref{fig:sentbytes} illustrates the number of MBytes sent for these approaches. For truncated mirroring, we have tested two scenarios. In the first scenario, we mirror only the TCP headers, in the second we mirror additional bytes to guarantee that the payload containing the http header is also sent to the analysis tool to allow for deep inspection as needed for measuring, e.g., success rate, as discussed in Section~\ref{MA}.

Figure~\ref{fig:sentbytes} shows selective forwarding and tunneling via vxlan send only 5\% respectively 6\% less bytes than port mirroring although they send 30\% less packets. The reason is that these approaches avoid  control messages that are typically small in size. Furthermore, vxlan adds some bytes to each packet for encapsulation (i.e. layer 3, 4 and vxlan headers) which leads to more bytes than selective mirroring.  The two truncated mirroring approaches send around 95\% fewer bytes compared to mirroring the whole packet because the TCP header is only about 5\% of the maximum segment size of TCP packets. Still, our OVS offsite analysis sniffer sends the least amount of bytes, only  1.25\% of what port mirroring sends. The reason is that it is customized and extracts only the target fields to be sent to the analysis node. 

\subsection{Summary}
Conducting complex application-independent analysis on the end host, as done by Tshark, can have a considerable negative impact on application performance. Network-based application monitoring approaches decouple log collection and analysis from the end components, allow for flexible placement of data collection and even analysis somewhere in the network.

Port mirroring has shown better performance than
tunneling in our experiments as it introduces less overhead
in the switch. However, port mirroring produces more traffic, which may negatively affect the overall cloud performance should the analysis
tool resides on a different node than the switch. Port mirroring also boosts the load on the analysis tool by increasing the number of messages that have to be processed; this is quite evident by the growth in of CPU utilization. Selective mirroring and tunneling reduces slightly this communication overhead. Truncated mirroring has the least communication overhead between mirroring approaches. However, it has significant impact of the switch latency and CPU utilization. 

Compared to mirroring and tunneling, the OVS sniffer has the advantages that: (a) it does not introduce any direct delay at the switch, and (b) it can perform some analysis locally or send selective information for remote analysis which significantly reduces the communication overhead. The disadvantage of the OVS sniffer is that it can be only implemented in software and not on SDN-enabled hardware switches. 

We believe that a network approach provides us
considerably more flexibility in the placement of monitoring
functionality. For hardware switches, selective mirroring is probably the most efficient approach if the analysis tool resides on a node that has a direct link to this switch, but tunneling provides flexibility for location of analysis tool that is probably worth the overhead. For software switches, we believe that our customized OVS sniffer is the preferred route to go because of performance and flexibility. 

\section{Related work} \label{related}
Monitoring has always been a fundamental aspect of distributed systems and large-scale networks and the literature is vaste. Yet, to the best of our knowledge none of the existing work exploits the functionality of SDN to enable effective and efficient application-level monitoring. 
\paragraph{Application monitoring via instrumentation:}Many
frameworks use \emph{instrumentation} \cite{twitter, twitter2, SAAD, PivotTracing, LEARNINGTOLOG, twitter3,semanticlogging}. For instance, Twitter \cite{twitter} instruments its code to generate structured \emph{client event} log messages. The log messages keep track of session information such as the client ID, IP address and request receiving time. The generated log messages can be later used for application performance measures such as request service time. Another example is SAAD \cite{SAAD}, which uses log statements as trace points to track
the execution flow of tasks during run-time, and exploits
the statistical similarity of tasks to detect performance anomalies. However, the use of instrumentation and application-dependent logs for analysis relies on experts with deep knowledge of the system, and can not be generally applied in larger settings. 
\paragraph{Application monitoring using SDN:}
There has been little work that exploits network functionality for application monitoring. NetAlytics \cite{Timothy2016} is deemed the closest to our work. NetAlytics deploys analysis nodes, called monitoring agents,  across the cloud network and connects them directly to the TOR SDN switches. These switches then apply port mirroring to collect data for conducting real-time analysis at the monitoring agents.  The main focus is on analyzing traffic that arrives at the monitor in real-time by developing specific parsers and a query language that allows to specify which flows need to be monitored and how. In this paper, we have a closer look at the forwarding mechanisms in the network themselves as well as analyze the option of conducting some of the analysis at the switch using an OVS sniffer. 
\paragraph{Network monitoring using SDN:}
Network monitoring using SDN has been extensively studied ~\cite{DAAS, SDN-PANDA, portmirroring, SPI, selectivemirror}. Everflow \cite{selectivemirror} utilizes the \emph{match and mirror} capability in commodity switches  to capture certain packets for network monitoring purpose (debug DCN faults). In this paper, we investigated the possibility and effectiveness of adapting the same approach but for application-level performance metrics such as HTTP request service time. Selective tunneling has been one of several approaches analyzed earlier in this paper in the context of application monitoring. 

Some approaches use port sniffing for monitoring purposes ~\cite{ Appflowsatruntime, Timothy2016, WAP5, blackblox, NetCheck}. For example, NefPerf \cite{nevperf} sniffs packets on all communication paths between NFVs in NFV deployments to compute per-hop throughput and delays, and uses
these measurements to identify both hardware and software performance bottlenecks. Our port sniffer has the same goals but with more focus on  enabling application monitoring and supporting metrics such as the ones described in Section~\ref{MA}. 
\paragraph{Switch software enhancements:} A further alternative to extend monitoring in the switch is to extend the switch software to include advanced analysis functionality \cite{DPI, UMON, Ringbuffer, ApplicationAware, UDA, UDA2}. However, performance evaluations of these approaches  have shown considerable overhead and impact on the performance of the switch's main task -- packet forwarding. Also, this requires extensions to the switch's code base. Thus, there are concerns on the adoption of these approaches in practice. 
Specifically, the authors of \cite{DPI} extend the OpenFlow architecture to be able to inspect the payload of the packets by inserting a set of predefined string patterns into the switch. Mekky et al. \cite{ApplicationAware} augment SDN switches with application processing logic defined in a table called application table. This enables customized packet handling in the
SDN data plane switch. Both approaches can handle significantly less packets per
time unit than standard switches. The approach of both \cite{UMON} and \cite{Ringbuffer} is to modify the OVS source code in order to decouple the monitoring functions from the forwarding path of OVS. UMON \cite{UMON} decouples monitoring from forwarding by defining
a monitoring flow table in the OpenFlow switch user space to separate monitoring rules from forwarding
rules. Zha el al. \cite{Ringbuffer} extends the kernel space of OVS to buffer the monitored packets into a ring buffer to be picked up by the monitoring process. However, both approaches are meant for network layer measurements and do not support any application level measurements.
\section{Conclusions and future work} \label{Conclusion}
 In this paper, we have explored the various trade-offs of different host and network based logging of monitoring data. In addition, we have proposed and implemented a network-based logging approach and provided a quantitative comparison.  The results show that sniffer tools on end-hosts can create considerable overhead. In contrast network-based approaches enable a flexible deployment and analysis. The use of SDN keeps the overhead induced by the data collection process at low levels. Virtualized switches further enhance the possibilities of what can be done at the network components. 

As a next step, we will evaluate architectures with top-rack OVS based switches that build on fast packet processing libraries. Furthermore, we will explore further high-level application monitoring functions to build a complete Monitoring-as-a-Service of the network.  

\bibliographystyle{ACM-Reference-Format}
\bibliography{sample-base}

\end{document}